\documentclass[prb,showpacs,twocolumn]{revtex4}
\usepackage{graphicx}
\usepackage{dcolumn}
\usepackage{bm}

\begin{document}

\preprint{}

\title{Graphene superlattice with periodically modulated Dirac gap}

\author{G.M. Maksimova, E.S. Azarova, A.V. Telezhnikov, and V.A. Burdov}

\affiliation{Department of Theoretical Physics, University of Nizhny Novgorod,
              23 Gagarin Avenue, 603950 Nizhny Novgorod, Russian Federation}

\date{\today}

\begin{abstract}

Graphene-based superlattice (SL) formed by a periodic gap
modulation is studied theoretically using a Dirac-type
Hamiltonian. Analyzing the dispersion relation we have found that
new Dirac points arise in the electronic spectrum under certain
conditions. As a result, the gap between conduction and valence
minibands disappears. The expressions for the positions of these
Dirac points in ${\bf k}$-space and threshold value of the
potential for their emergence were obtained. Also, the dispersion law and renormalized group velocities around the new Dirac points were calculated. At some parameters of
the system, we have revealed interface states which form the top
of the valence miniband.

\end{abstract}


\maketitle

\section{Introduction}

During for the last years extremely much attention was paid to the
electronic properties of graphene (see Ref.1 for the review). Such
interest results, in particular, from the fact that physics of the
low-energy carriers in graphene is governed by a Dirac-type
Hamiltonian. The band structure of an ideal graphene sheet has no
energy gap. As a consequence, Dirac electrons become massless, and
reveal unusual properties such as perfect transmission at normal
incidence through any potential barrier (the Klein
paradox\cite{{Klein,Kat}}), trembling motion (or
Zitterbewegung\cite{GM}), etc. Because of the Klein effect, an
electrostatic potential cannot confine electrons in graphene. This
property of graphene impedes its use in electronic
devices.\cite{Geim} However, as has been shown,\cite{Martino} it
is possible to confine massless Dirac particle in graphene sheet
by inhomogeneous magnetic field. The confinement can be also
achieved by combining electric and uniform magnetic
fields.\cite{N7,N8} Meanwhile, Dirac electrons can be localized
electrostatically in a gapped graphene.

The gap can be induced by substrate or strain engineering as well as by
deposition or adsorption of molecules on a graphene layer. For
example, two carbon sublattices of graphene placed on top of
hexagonal boron nitride ($h$-BN) become nonequivalent due to their
interaction with the substrate. The band-structure calculations
within the local-density approximation for this system gives a gap
not less than 53 meV.\cite{Gio} A hydrogenated sheet of graphene
(graphane) is a semiconductor with a gap of the order of a few
eV.\cite{Leb}

Besides, it is possible to modulate spatially the gap (i.e. the
particle's mass) in graphene. It was shown that the spatial mass
dependence leads to suppression of Klein tunneling and induces
confined states.\cite{{Peres,Nori}} The required gap modulation
can be created, for instance, in graphene placed on a substrate
fabricated from different dielectrics. It is also possible to use
for this purpose an inhomogeneously hydrogenated graphene or
graphene sheet with nonuniformly deposited CrO$_{3}$ molecules.
Correspondingly, one can fabricate different graphene
heterostructures with the gap discontinuity. In particular,
graphene-based superlattice (SL) can be formed due to the periodic
modulation of the band gap.

Recently, electronic structure of graphene under external
periodic potential has been the subject of numerous
studies.\cite{B1,B2,Yang,Fertig,Vas2,B6,B7,B8,B9,B10,B11,I1,I2,I3,I4,I5}
Increasing interest in graphene SLs results from the prediction of
possible engineering the system band structure by the periodic
potential, which opens new ways for fabrication of graphene-based
electronic devices. The graphene SLs have been realized
experimentally. For example, graphenes grown epitaxially on metal
surfaces\cite{I1,I2,I3,I4} demonstrate SL patterns with about
several nanometers SL period. Recent scanning tunneling microscopy
studies\cite{I5} of corrugated graphene monolayer on Rh foil show
that the quasi-periodic ripples generate a weak one-dimensional
electronic potential in graphene leading to emergence of the SL
Dirac points. It was as well shown theoretically that a
one-dimensional periodic potential really affects the transport
properties of graphene. For instance, the Kronig-Penney (KP)-type
electrostatic potential induces strong anisotropy in the carrier
group velocity around the Dirac point\cite{B1,Vas2} leading to the
so-called supercollimation phenomenon.\cite{B1,B2} Besides, in the
SL spectrum new (extra) Dirac points appear in Brillouin zone.
These features has been also examined for the different types of
graphene SL including the magnetic KP-SL with delta-function
magnetic barriers.\cite{B6,B7,B8,B9,B10,B11} In Ref.19 the
first-principles studies of the electronic structure of
graphene-graphene SL modeled with a repeated structure of pure and
hydrogenated graphene (i.e., graphane) strips has been performed.
It was found that unlike other graphene nanostructures, the
hydrogenated graphene SLs exhibit both direct and indirect band
gaps.

In this paper we focus on the electronic states in graphene-based
SL with periodically modulated gap and relative band shift
(potential) where the gap and potential are piecewise constant
functions of $x$. The model of such SL has been considered earlier.\cite{Rat} However, no detailed examination of the electronic structure in such type SLs in a wide range of the
system parameters has been carried out. In particular, we have
first found that the considered SL can be gapped or gapless
depending on the band shift. We analyze in detail the SL with
equal widths of the gapless and gapped graphene fractions, and
show that the forbidden miniband exists up to some threshold value
$V=V_c$ corresponding to the first emergence of the Dirac-like
point at ${\bf k}=0$. When the potential $V$ exceeds $V_c$, this
Dirac-like point disappears, opening the minigap at ${\bf k}=0$.
At the same time, two extra Dirac points arise in symmetric
positions on the $k_y$ axis. This scenario differs from the one
realized in graphene SL formed by the electrostatic
potential,\cite{Vas2} where the original Dirac point (at ${\bf
k}=0$) always exists. Further $V$ increasing leads to emergence of
a new pair of the Dirac points from the origin ${\bf k}=0$. The
dispersion in the vicinity of these points appears to be
anisotropic, indicating an anisotropic renormalization of the
group velocity. We also show that interface states can exist in
the gap-induced SL in contrast to the SL formed by a
Kronig-Penney-type electrostatic potential. For this case the
relevant conditions and values of the system parameters were
obtained.

\section{Dirac particle in graphene superlattice}

Let us consider one-dimensional (1D) graphene superlattice with
period $l$ formed by position-dependent gap and band shift. As was
shown in Ref.29, such structure can be realized, e.g., on the base
of graphene deposited on a strip substrate combined from silicon
oxide and $h$-BN (Fig.1). The SL electronic structure in the
vicinity of ${\bf K}$-point of the Brillouin zone is described by
the Dirac-like Hamiltonian
\begin{eqnarray}
{\hat H}=\upsilon_{F}{\hat{\bf
p}}\boldsymbol{\hat{\sigma}}+V(x){\hat
1}+\Delta(x){\hat\sigma_{z}}
\end{eqnarray}
where ${\hat{\bf p}}$ is the momentum operator, ${\hat\sigma}_i$
are the Pauli matrices, ${\hat 1}$ is a unit $2\times 2$ matrix,
$\upsilon_{F}$ is the Fermi velocity, and $\Delta(x)$, $V(x)$ are
periodic functions equal to $\Delta$ and $V$, respectively, at
$a\leq x\leq l$, and zero at $0\leq x<a$. Here, the potential $V$
defines the shift of the forbidden band center in the gapped
graphene with respect to the Dirac point in the gapless
graphene\cite{Rat,Silin} (see Fig.1). Generally speaking, the
Fermi velocity can differ in graphene modifications placed on
different substrates. In our model, however, we neglect the
dependence $\upsilon_{F}$ on $x$ supposing
$\upsilon_{F}\approx10^{8}$ cm/s in both graphene fractions.

The Kronig-Penney model considered here as applied to graphene SL is also used in many other physical problems including, e.g., modeling of semiconductor SLs or relativistic particle dynamics. Although master equations describing the single-particle evolution differ for all the mentioned systems, explicit forms of the dispersion relations obtained within the framework of the Kronig-Penney model resemble to each other. Nevertheless, individual features of these systems result in qualitative distinctions in their band structures. For instance, the dispersion relation for 1D relativistic electron in Kronig-Penney potential\cite{Strange,Kel} is similar to the one obtained in this paper (see Eq.(9) below) for 2D electrons in 1D graphene SL. In our case, however, the periodic potential in Dirac-like equation is formed by two terms having different meanings from the point of view of relativistic physics. The relative band shift $V$ can be treated as the time-like vector component while the band-gap $\Delta$ represents a scalar potential. This circumstance together with the two-dimensionality of the electron gas provide new fundamental properties of the SL electronic structure (such as appearance of extra Dirac points), which will be discussed in detail in section III.

The Dirac equation
\begin{eqnarray}
{\hat H}\Psi(x,y)=E\Psi(x,y)
\end{eqnarray}
admits the solutions $\Psi(x,y)=\exp(ik_{y}y)\Psi(x)$,
where the two-component spinor envelope function $\Psi(x)$
satisfies the equation
\begin{eqnarray}
i\frac{d\Psi}{dx}={\hat h}(x)\Psi(x)
\end{eqnarray} with
\begin{eqnarray}
{\hat h}(x)=\pmatrix{ik_{y} &
\frac{V(x)-E-\Delta(x)}{\hbar\upsilon_{F}}\cr
\frac{V(x)-E+\Delta(x)}{\hbar\upsilon_{F}} & -ik_{y}\cr}.
\end{eqnarray}

\begin{figure}[tbp] \centering
\includegraphics*[scale=0.47]{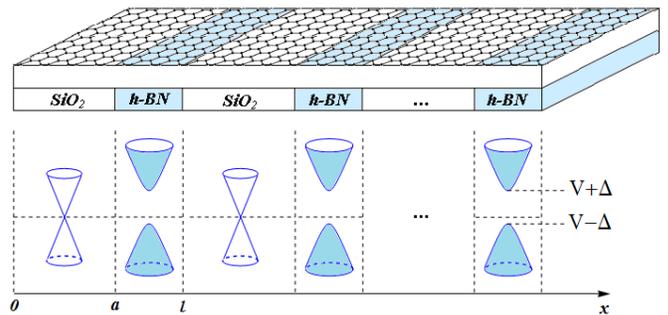}
\caption{(Color online) Top: graphene layer on the striped
substrate composed of silicon oxide and hexagonal boron nitride.
Bottom: schematic diagram showing the electronic energy spectrum
in graphene SL.} \label{figure1}
\end{figure}

The formal solution of Eq.(3) is
\begin{eqnarray}
\Psi(x)={\hat\Re}_{x}\exp\left(-i \int_{x_{0}}^{x}{\hat
h}(x_{1})dx_{1}\right)\Psi(x_{0}),
\end{eqnarray}
where
${\hat\Re}_{x}$ is the spatial ordering operator.\cite{Kel,Ar}
This expression can be simplified as
\begin{eqnarray}
\Psi(x)=\exp(-i(x-x_{0}){\hat h})\Psi(x_{0})
\end{eqnarray}
if both
points $x$ and $x_{0}$ belong to the space-homogeneous region.

In this case it is convenient to define the matrix
\begin{eqnarray}
{\hat t}(x-x_{0})=\exp(-i(x-x_{0}){\hat h}).
\end{eqnarray}
Here, the matrix ${\hat h}$ is defined by Eq.(4) where
$V(x)=\Delta (x)=0$ if $0\leq x, \ x_0<a$, and $V(x)=V$, $\Delta
(x)=\Delta$ if $a\leq x, \ x_0\leq l$. The straightforward
calculation yields ${\hat h}^2(x)=K^2(x){\hat 1}$, where
$K(x)=\sqrt{\big((V(x)-E)^2-\Delta^2(x)\big)/(\hbar\upsilon_{F})^2-k_{y}^2}$.
Therefore, all even powers in the Taylor series of the exponential
function in Eq.(7) will be proportional to the unit matrix while
all odd powers will be proportional to the matrix ${\hat h}$
itself. This leads to the following expression:
\begin{eqnarray}
&&{\hat t}(x-x_{0})={\hat 1}\cos\alpha -
i{\hat h}(x){\sin\alpha\over K(x)}= \nonumber \\
&&\pmatrix{\cos\alpha+\frac{k_{y}}{K(x)}\sin\alpha &
-i\sin\alpha\frac{V(x)-E-\Delta(x)}{\hbar\upsilon_{F}K(x)}\cr
-i\sin\alpha\frac{V(x)-E+\Delta(x)}{\hbar\upsilon_{F}K(x)} &
\cos\alpha-\frac{k_{y}}{K(x)}\sin\alpha\cr},~~~
\end{eqnarray}
where $\alpha=(x-x_{0})K(x)$.

We can now find that $\Psi(l)={\hat T}\Psi(0)$, where ${\hat
T}={\hat t}(l-a){\hat t}(a)$. Note, that $\det(t)=1$, and,
consequently, $\det(T)=1$ as well. This equality and the Bloch
condition $\Psi(l)=\Psi(0)\exp(ikl)$ (here, $k$ is the Bloch wave
vector) yield the dispersion relation $2\cos(kl)=Tr(T)$ for the 1D
graphene-based SL. Using Eq.(8) one can find
$Tr(T)=Tr(t(l-a)t(a))$. Accordingly, the resulting dispersion
equation reads
\begin{eqnarray}
&&\cos(kl)=\cos(k_{x}a)\cos\left(q_{x}(l-a)\right)+ \nonumber \\
&&\frac{EV-(\hbar\upsilon_{F}k_{x})^2}{(\hbar\upsilon_{F})^2
k_{x}q_{x}}\sin(k_{x}a)\sin\left(q_{x}(l-a)\right),
\end{eqnarray}
with $k_{x}\equiv K(+0)$, $q_{x}\equiv K(l-0)$. At
$(V-E)^2-\Delta^2<(\hbar\upsilon_{F}k_{y})^2$, the wave number
$q_{x}$ is imaginary and the allowed energies are given by Eq.(9)
with $q_{x}$ replaced by $i|q_{x}|$.

Eq.(9) for the allowed energies was obtained also in Ref.29 via
wave function matching. However some results of this work
concerning the miniband structure and existence of the interface
states seem questionable. Below we discuss these issues in detail.
Note that at $\Delta=0$, Eq.(9) coincides with the one found for
single-layer graphene in a periodic piecewise constant potential
$V(x)$.\cite{Ar,Vas1,Vas2}

\section{Electronic structure}

The miniband structure depends on the system parameters as well as
on the $y$-component of the wave vector. As seen from the Eq.(9),
the dispersion relation is invariant with respect to the
simultaneous replacements $E\rightarrow -E$, $V\rightarrow -V$.
Therefore, in what follows we shall consider, for definiteness,
only nonnegative values of the relative band shift: $V\geq 0$.
When $V=0$, the energy spectrum is completely symmetric related to
the value $E=0$ corresponding to the original Dirac point in the
gapless graphene. In this case the two first minibands
symmetrically situated above and below the point $E=0$
are the conduction and valence ones, respectively. As $V$
increases, the conduction and valence minibands gradually shift
up. Below we shall concentrate on these two minibands only,
assuming the Fermi level to be in between at any $V$.

The electron and hole energies as functions of $k$ at $k_{y}=0$,
$\Delta=26.5$ meV for different values of the potential $V$ and
widths $a$ are plotted in Fig.2(a). The parameters we choose are
appropriate for graphene. Thick and thin solid lines correspond to
$V=0$ and $V=24.2$ meV, respectively, and $a=l/2$. When $V$
becomes nonzero the electron (conduction) and hole (valence)
minibands shift up, and the electron miniband turns out to be a
little narrower than the hole one. This is, presumably, due to the
fact that at chosen values of the parameters the electron miniband
completely forms under the barrier at $a<x<l$:
$V-\Delta<E<V+\Delta$, while the hole-miniband energies mainly
belong to the over-barrier region $E<V-\Delta$.

The energy branches in the conduction and valence minibands at $a=l/6$, $V=24.2$ meV is shown with dashed line. We can
see that, as the gapped graphene fraction in the SL increases, the
electron-hole minigap increases too. Nevertheless, in any case, at
$k_{y}=0$ the minigap cannot exceed the gap value
$2\Delta$.\cite{text} Indeed, it is clear that if the whole
graphene layer is gapped, i.e. $a=0$, the forbidden miniband
should be $2\Delta$ (at $k_{y}$=0).

\begin{figure}[tbp] \centering
\includegraphics*[scale=0.6]{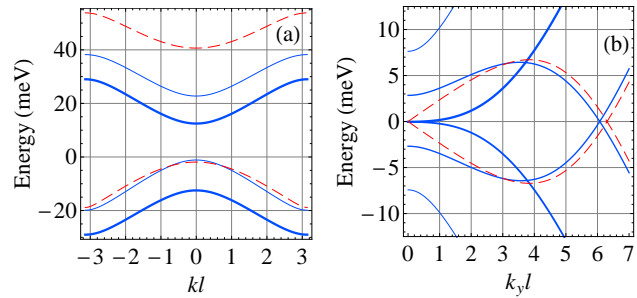}
\caption{(Color online) (a). Two low-energy minibands of the SL
spectrum with $k_{y}=0$, $l=60$ nm, and $\Delta=26.5$ meV. Thick
(blue) line corresponds to the band shift $V=0$, and
$a=l/2$. Thin (blue) and dashed (red) lines correspond to the
widths of gapless graphene fraction $a=l/2$ and $a=l/6$, respectively, at
$V=24.2$ meV. (b). $k_{y}$-dependence of the electron
energies in the conduction and valence minibands at $k=0$, $l=60$
nm, $a=l/2$, $\Delta=0$, and $V=195.7$ meV (dashed red line). For other
curves $\Delta=26.5$ meV and $V=77.1$ meV (thin blue solid line);
$V=143.26$ meV (thick blue solid line),
and $V=195.7$ meV (solid blue line). The zeroth energy corresponds
to the minigap center for thin solid line, and to the contact and cone-like Dirac
points for all the others.} \label{figure2}
\end{figure}

Figure 2(b) illustrates the dependence of the electron and hole
energies on $k_y$ at different $V$ and $\Delta$, and $k=0$,
$a=l/2$, $l=60$ nm. We show only semiaxis $k_y>0$ because of the
symmetry $k_y\rightarrow -k_y$ in the dispersion law [Eq.(9)]. The
energies in the figure are counted from the minigap center (its
position determined at $k=k_y=0$ depends on $V$) for $V=77.1$ meV
(thin solid line) at
$\Delta=26.5$ meV. In other cases the energy origin coincides with
the contact or cone-like point energy. Our calculations demonstrate a
monotonous expansion of the forbidden miniband with $|k_{y}|$
increasing (thin solid line in Fig.2(b)) until
$V$ becomes greater than some threshold value $V_c$ (for chosen
set of parameters $V_c=143.26$ meV). When $V=V_c$ (thick solid
line in Fig.2(b)), the electron and hole energy branches touch each
other at $k=k_y=0$ closing the minigap. When $V$ exceeds $V_c$,
the minigap at $k=k_y=0$ opens, but two extra Dirac points appear
in symmetric positions on the $k_y$-axis (solid line in
Fig.2(b)), and never then vanish. Thus, beginning with $V=V_c$ the gap-induced SL becomes gapless.

This feature of the theoretically predicted dependence of the SL
minigap on $V$, in fact, can be observed by spectroscopic methods.
In order to vary the parameter $V$ it is possible to apply
additional nano-gate potential to the gapped graphene fractions in
the SL. Closing the minigap accompanied by the disappearance of
the absorption edge and appearance of the contact point in the SL
spectrum allows one, in turn, to determine experimentally the
threshold value $V=V_c$.

Further increase of $V$ results in formation of new additional
cone-like Dirac points which originate from $k_y=0$. Such a
behavior is not unique. As was shown in Refs 15-17 for gapless
graphene, the new Dirac points can emerge in the presence of a
sinusoidal or squarewell SL potential. A simple analytical
expression for the positions of these points in ${\bf k}$-space
has been obtained by different research groups.\cite{Vas2,Ar} In
our model, however, in contrast to the SLs discussed in the works
quoted above, the Dirac point being a prototype of the original
Dirac point at ${\bf k}=0$ arises at certain finite values of $V$
equal to $V_n=2\pi n\hbar\upsilon_F/l+\sqrt{(2\pi
n\hbar\upsilon_F/l)^2+\Delta^2}$ (defined by the zeroes of
$k_{yn}$, see Eq.(12) below) with $n$ being positive integer. Note
that, $V_1$ coincides with the threshold potential value $V_c$.

It is clearly seen from Fig.2(b) that extra Dirac point located at
$k_{y}l\simeq 6.3$ for $\Delta=0$ (dashed line) remains also in
the case $\Delta\ne 0$ at the same values of $V$ when $\Delta\ll
V$. However, the location of the Dirac point in this case slightly
shifts towards zero with $\Delta$ increasing. In order to find the
location of the Dirac points in ${\bf k}$-space we note that
$k_{x}(E,k_{y})$ and $q_{x}(E,k_{y})$ coincide at
\begin{eqnarray}
E_0(V)={V^2-\Delta^2\over 2V}.
\end{eqnarray}
Correspondingly, Eq.(9) at $a=l/2$, $k=0$ and $E=E_0$ turns into
\begin{eqnarray}
1=\cos^2\frac{k_{x}l}{2}+\frac{E_0V-E_0^2+\ \hbar^2\upsilon_{F}^2
k_{y}^2}{(\hbar\upsilon_F k_{x})^2}\sin^2\frac{k_{x}l}{2}.
\end{eqnarray}
Obviously this equation is satisfied if $k_{x}l=2\pi n$
($n$ is integer and differs from zero), which leads to
$k_y=k_{yn}$, where
\begin{eqnarray}
k_{yn}={1\over l}\sqrt{\left(\frac{E_0l}{\hbar\upsilon_{F}}\right)^2-(2 \pi
n)^2}.
\end{eqnarray}

One can check that for $a=l/2$, the right side of Eq.(9) has local
maximum (equal to unity) at $E=E_0$ and $k_y=k_{yn}$. This means
coincidence of conduction and valence miniband edges.

According to Eq.(12), in general case, when the ratio
$E_{0}l/2\pi\hbar\upsilon_{F}$ is not integer, the number of Dirac
points $N_{D}$ symmetrically situated around $k_{y}=0$ is given by
\begin{eqnarray}
N_{D}=2\left[\frac{E_0l}{2\pi\hbar\upsilon_{F}}\right],
\end{eqnarray}
where $[...]$ denotes an integer part. One of the contact points
is always located at $k_{y}=0$ if $E_0l/\hbar\upsilon_{F}=2\pi n$.
In this case $V=V_n$, and the total number of Dirac points is odd:
$N_{D}=2n-1$.

Thus, now we can write analytically the inequality for the system
parameters corresponding to the formation of the gapless SL.
Evidently, the gap separating the conduction and valence minibands
vanishes if
\begin{eqnarray}
\frac{(V^2-\Delta^2)l}{4\pi\hbar\upsilon_{F}V}\geq 1.
\end{eqnarray}
The case of the rigorous equality in Eq.(14) corresponds to the
threshold value of the potential:
$V_c=V_1=2\pi\hbar\upsilon_F/l+\sqrt{(2\pi\hbar\upsilon_F/l)^2+\Delta^2}$
(the positive root of the quadratic equation).

\begin{figure}[tbp] \centering
\includegraphics*[scale=0.47]{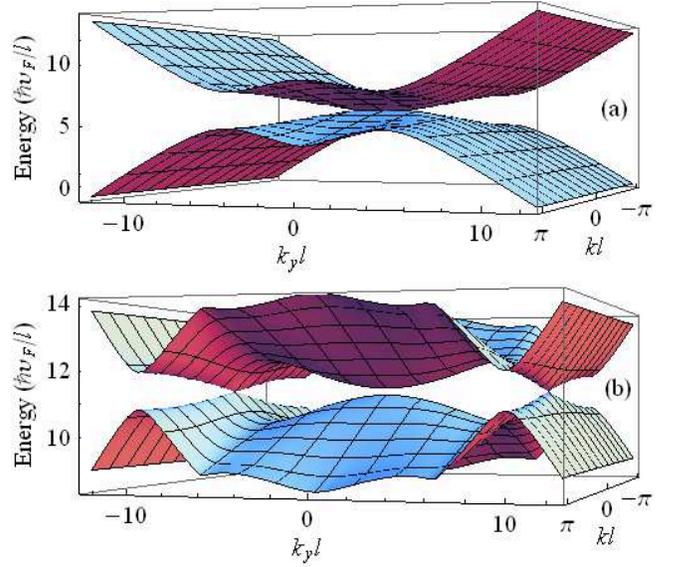}
\caption{(Color online) Energy surfaces near the: contact point $k=k_y=0$ at $V=V_c=143.26$ meV (a); cone-like Dirac point $k=0$, $k_y=k_{y1}$ at $V=165.15$ meV (b). $\Delta=26.5$ meV.} \label{figure3}
\end{figure}

Solving Eq.(9) numerically we have found the dispersion law
presented in Fig.3 for $V=V_1=V_c$ (panel (a)) and $V_1<V<V_2$
(panel (b)). It is seen that when $V=V_c$ there is the only
contact point in the electronic spectrum, having no conical shape.
When $V>V_c$, two cone-like Dirac points exist at $E=E_0(V)$,
$k=0$, and $k_y=\pm k_{y1}$. Let us further consider the
dispersion relation in the nearest vicinity of the contact and
cone-like Dirac points.

We start with the contact point arising at $E=E_0(V_n)$, $k=0$,
and $k_y=0$. For this purpose one has to expand the terms in
Eq.(9) at $V=V_n$ into the Taylor series up to the lowest powers
of $E-E_0(V_n)$, $k$, and $k_y^2$. As an example, below we
consider the case $n=1$, i.e. $V=V_c$. After some algebra one
obtains the dispersion relation in the following form:
\begin{eqnarray}
&&\varepsilon={d^4(k_yl)^2\over 4\pi u_c^2(1+3d^2)}\pm \nonumber \\
&&\sqrt{{(kl)^2\over 1+3d^2}+{1+3d^2+d^6/u_c^3\over
(1+3d^2)^2}{d^2(k_yl)^4\over (4\pi)^2u_c}},
\end{eqnarray}
where we have introduced dimensionless energy $\varepsilon=(E-E_0)l/\hbar\upsilon_F$, and parameters $u_c=lV_c/4\pi\hbar\upsilon_F$ and
$d=l\Delta/4\pi\hbar\upsilon_F$. The signs "+" and "$-$"
correspond to the electrons and holes, respectively. Evidently,
the energy surface $\varepsilon(k,k_y)$ around the only contact
point, indeed, has no conical form. The first term in the right
side of Eq.(15) defines an asymmetry of the conduction and valence
minibands related to $\varepsilon=0$. If $d\rightarrow 0$ (in this
case $u_c\rightarrow 1$), the conduction and valence minibands
become symmetric near the contact point. According to Eq.(15) at
$k=0$, the edges of the conduction and valence minibands are
parabolic along $k_y$-axis:
\begin{eqnarray}
\varepsilon\sim d\left({d^3\over
u_c^{3/2}}\pm\sqrt{1+3d^2+{d^6\over u_c^3}}\right)k_y^2.
\end{eqnarray}
As a result, the dispersion along $k_y$ is almost flat. In the limiting case $d\ll 1$, the coefficient before $k_y^2$ becomes proportional to $d$ (i.e. $\Delta$). Note that in the SL based on a gapless graphene the dispersion curves at $k=0$ also demonstrate a nonlinear behavior, where, however, dispersion is defined by the third power of the $y$-component of the wave vector: $\varepsilon\sim\pm k_{y}^3$.\cite{Vas2}

\begin{figure}[tbp] \centering
\includegraphics*[scale=0.68]{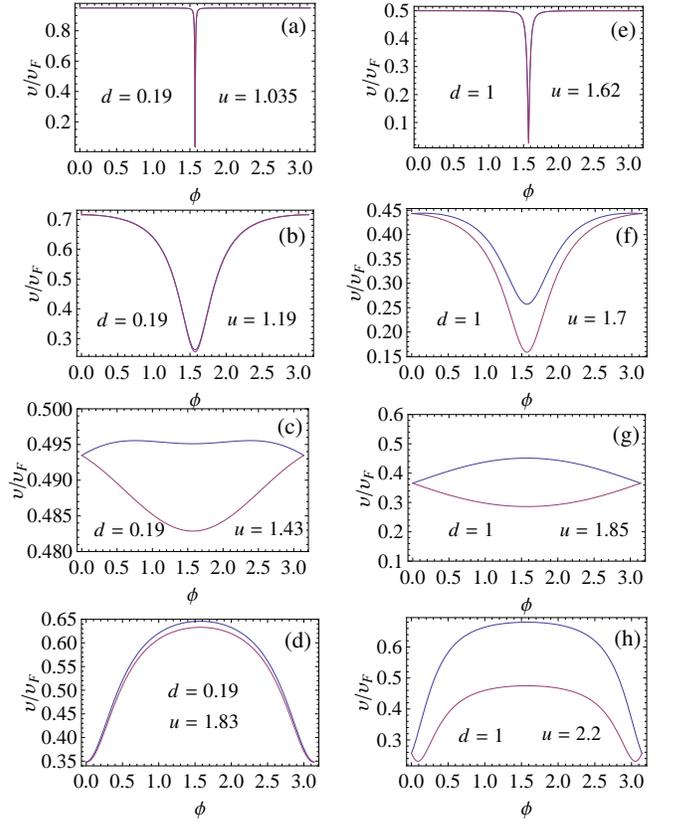}
\caption{(Color online) Angle dependence of the absolute value of
the electron and hole (upper and lower curves, respectively, in
each panel) velocities normalized to $\upsilon_F$ near the contact
point $k=k_y=0$ at $Ql=0.5$, and $V=V_c$ (panels (a) and (e)), and near the cone-like Dirac point $k=0$, $k_y=k_{y1}$ (panels (b) - (d) and (f) - (h)). Values of $d$ and $u$ are indicated in the figure.}
\label{figure4}
\end{figure}

Basing on Eq.(15) we can now determine the velocity components
$\upsilon_k=\hbar^{-1}\partial E/\partial k$,
$\upsilon_y=\hbar^{-1}\partial E/\partial k_y$ in the
vicinity of the contact point at $V=V_c$. Calculating the
derivatives one gets
\begin{eqnarray}
{\upsilon_k\over\upsilon_F}=\pm{k\over\sqrt{(1+3d^2)k^2+\left(1+3d^2+{d^6\over
u_c^3}\right){l^2d^2k_y^4\over (4\pi)^2u_c}}},
\end{eqnarray}
\begin{eqnarray}
&&{\upsilon_y\over\upsilon_F}={ld^4k_y\over 2\pi u_c^2(1+3d^2)}\pm \nonumber \\
&&{\left(1+{d^6\over (1+3d^2)u_c^3}\right){2l^2d^2\over
(4\pi)^2u_c}k_y^3\over\sqrt{(1+3d^2)k^2+\left(1+3d^2+{d^6\over
u_c^3}\right){l^2d^2k_y^4\over (4\pi)^2u_c}}}.
\end{eqnarray}
The obtained expressions exhibit strong anisotropy of the electron
and hole velocities in $(k,k_y)$-plane, which is clearly seen from
the dependence of the absolute value
$\upsilon=\sqrt{\upsilon_k^2+\upsilon_y^2}$ on the polar angle
$\phi$ introduced with standard relations: $k=Q\cos\phi$,
$k_y=Q\sin\phi$, where $Q=\sqrt{k^2+k_y^2}$. The dependence
$\upsilon (\phi)$ is sensitive to the induced gap $\Delta$ (or
$d$) as Fig.4 shows. When $d\ll 1$ (panel (a)), $\upsilon$ is
close to $\upsilon_F$ for all angles except for the nearest
vicinity of $\pi/2$ where narrow dip in the dependence
$\upsilon(\phi)$ occurs. As $d$ increases, $\upsilon$ monotonously
decreases in whole, keeping the dip that becomes a little wider as
seen in panel (e). Such a dip is caused by the lens-like shape of
the energy surfaces $\varepsilon(k,k_y)$ both for the electrons
and holes as described by Eq.(15). In this case, the component
$\upsilon_k$ is much greater than $\upsilon_y$ for all $\phi$
except for the narrow domain near the "lens" edge, where $\phi$ is
close to $\pi/2$. Within this domain both $\upsilon_k$ and
$\upsilon_y$ are small as follows from Eqs (17) and (18). Note
also that, the electron and hole velocities $\upsilon_e$ and
$\upsilon_h$, in fact, coincide at the considered values of $d$.

Finalizing discussion of the energies and group velocities near
the contact point we would like to emphasize as well that,
expressions similar to Eqs (15) - (18) can be easily obtained for
$V=V_n$. In this case one should simply replace $\pi$ by $\pi n$
in all the expressions. Consequently, the dispersion in $k_y$
direction flattens compared to the one described by the
Eq.(16).

When $E_0l/\hbar\upsilon_{F}>2\pi n$, which implies that $V>V_n$,
the situation is different. In this case introducing small
deviation $\tilde{k_y}=k_y-k_{yn}$, it is easy to find from Eq.(9)
the dispersion law $\varepsilon (k,\tilde{k_y})$ in the form:
\begin{eqnarray}
&&\varepsilon=\beta\tilde{k_y}l\pm\sqrt{\alpha^2(kl)^2+(\beta^2+
\gamma^2)({\tilde{k_y}}l)^2}.
\end{eqnarray}
Here, the parameters $\alpha$, $\beta$, and $\gamma$ are defined as
\begin{eqnarray}
&&\alpha={un^2\over\sqrt{(u^2-d^2)^2(u^2+d^2)+n^2d^4}}, \nonumber \\
&&\beta={d^2(u^2-d^2-n^2)\sqrt{(u^2-d^2)^2-u^2n^2}\over(u^2-d^2)^2(u^2+d^2)+n^2d^4},  \nonumber \\
&&\gamma=\sqrt{{\left((u^2-d^2)^2-u^2n^2\right)(u^2-d^2-n^2)\over(u^2-d^2)^2(u^2+d^2)+n^2d^4}},
\end{eqnarray}
where the dimensionless potential $u=lV/4\pi\hbar\upsilon_F$ is
always greater than $n$ if $\Delta\ne 0$.

Provided that $V$ slightly exceeds $V_n$ ($V-V_n\ll V_{n+1}-V_n$), the energy surface
becomes a cone elongated in $k_y$-direction. As $V$ increases, the
cone gradually turns into the isotropic one, and then becomes
elongated in $k$-direction. Thus, the dispersion law around the
new cone-like Dirac points in the considered SL has strong
anisotropy dependent on $V$ in contrast to the original Dirac
point in a gapless graphene. Similar behavior takes place in the
graphene SL formed by $V$-modulation at $\Delta=0$.\cite{Vas2} In
our case, however, in contrast to the mentioned work, the cone is
always tilted due to the presence of the linear in $\tilde{k_y}$
term in the dispersion relation [Eq.(19)]. The cone-axis obliquity
is defined by the coefficient $\beta$ which tends to zero when
$\Delta\rightarrow 0$.

The anisotropy of the dispersion law naturally manifests itself in the dependencies of the electron and hole velocities on $\phi$. Because of the cone-type shape of the energy surface, the velocity components
\begin{eqnarray}
&&{\upsilon_k\over\upsilon_F}=\pm{\alpha^2\cos\phi\over\sqrt{\alpha^2\cos^2\phi+(\beta^2+ \gamma^2)\sin^2\phi}}, \nonumber \\
&&{\upsilon_y\over\upsilon_F}=\beta\pm{(\beta^2+\gamma^2)\sin\phi\over\sqrt{\alpha^2\cos^2\phi+
(\beta^2+\gamma^2)\sin^2\phi}},
\end{eqnarray}
do not depends on $Q$ but exclusively on $\phi$. Due to the cone
tilt, $\upsilon_y$-component has nonzero mean value
\begin{eqnarray}
&&\langle\upsilon_y\rangle_{\phi}=\beta\upsilon_F
\end{eqnarray}
being the same for the electrons and holes.

In Fig.4 we have plotted the relation $\upsilon/\upsilon_F$ around the cone-like Dirac point with $k_y=k_{y1}$ for the electrons and holes as function of the polar angle $\phi$ for various $V>V_c(\Delta)$ ($u>u_c(d)$) at two values of the gap: $\Delta=26.5$ meV ($d=0.19$, panels (b) - (d)), and $\Delta=138.36$ meV ($d=1$, panels (f) - (h)). Provided that $d\ll 1$ the parameter $\beta$ is close to zero, and the Dirac-cone tilt is negligibly small. As a result, $\upsilon_e$ and $\upsilon_h$ differ insignificantly. The electron and hole velocities vary strongly when $\alpha^2$ strongly differ from $\beta^2+\gamma^2$ (panels (b) and (d), where $V$ equals 165.15 meV and 253.23 meV, respectively). This corresponds to the Dirac cone elongated in $k_y$-direction if $\alpha^2>\beta^2+\gamma^2$ (panel (b)) or in $k$-direction in the opposite case (panel (d)). When $\alpha^2\approx\beta^2+\gamma^2$ (this is so, e.g., at $V=196$ meV) the cone becomes almost isotropic. Consequently, $\upsilon_e$ and $\upsilon_h$ remain almost constant and equal to $\upsilon_F/2$ (panel (c)). At greater $d$ the cone tilt differently influences $\upsilon_e$ and $\upsilon_h$, as shown in panels (f) - (h), where $V$ equals 235.2 meV, 256 meV, and 304.4 meV, respectively. The behavior of the electron and hole velocities in this case is qualitatively similar to that takes place at $d\ll 1$. However the difference $\upsilon_e-\upsilon_h$ is no longer small.

We do not discuss here the case $a\neq l/2$. We expect that Dirac
points appear in this case too. However, the electron-hole
miniband profile should be significantly asymmetric related to the
Fermi level as it was in $V$-modulated SL.\cite{Vas2}

\section{Interface states}

As was shown by Ratnikov and Silin\cite{RS} interface states can
exist in graphene-based heterojunctions. It was found that
interface states result from the crossing of dispersion curves of
gapless and gapped graphene modifications. Meanwhile, in graphene-based superlattices, formation of interface minibands was considered to be impossible.\cite{Rat} In contrast to this statement we have found that the interface states can arise under certain
conditions discussed below. Since the wave functions of these
states behave as exponentials along $x$-axis, the wave vectors
$k_{x}$ and $q_{x}$ should be imaginary, so that
\begin{eqnarray}
(\hbar\upsilon_{F}k_{y})^2>E^2,~~~
(\hbar\upsilon_{F}k_{y})^2+\Delta^2>(E-V)^2.
\end{eqnarray}
In this case Eq.(9) transforms into
\begin{eqnarray}
&&\cos(kl)=\cosh(k'_{x}a)\cosh(q'_{x}(l-a))+ \nonumber \\
&&\frac{EV-E^2+(\hbar\upsilon_{F}k_{y})^2}{(\hbar\upsilon_{F})^2
k'_{x}q'_{x}}\sinh(k'_{x}a)\sinh(q'_{x}(l-a)),~~~~
\end{eqnarray}
with $k'_{x}=\sqrt{k_{y}^2-\frac{E^2}{(\hbar\upsilon_{F})^2}}$,
$q'_{x}=\sqrt{k_{y}^2+\frac{\Delta^2-
(V-E)^2}{(\hbar\upsilon_{F})^2}}$.

\begin{figure}[tbp] \centering
\includegraphics*[scale=0.65]{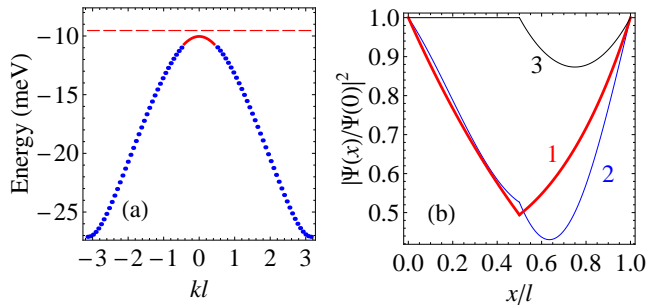}
\caption{(Color online) (a). Hole miniband of the SL spectrum for
$a=l/2=30$ nm, $V=11$ meV, $\Delta=22$ meV, and $k_yl=1$. Top of
the energy miniband depicted by the solid (red) line corresponds
to the interface states. Dashed (red) line shows the interface
energy level for an isolated heterojunction. (b). Probability
densities for: interface hole state (both $k'_x$ and $q'_x$ are
real) with $k_yl=1$, $kl=0.1$ (curve 1); oscillating hole states
with $k_yl=1$, $kl=2.5$ (both $k'_x$ and $q'_x$ are imagine ---
curve 2) and $k_yl=0$, $kl=0.4$ ($k'_x$ is imagine, $q'_x$ is real
--- curve 3). The SL spectrum parameters are the same to those
indicated in panel (a).} \label{figure5}
\end{figure}

Evidently the solution of this equation can exist only if
\begin{eqnarray}
(\hbar\upsilon_{F}k_{y})^2-E^2<-EV.
\end{eqnarray}
Since the left side of this expression is positive (see Eq.(23))
the allowed values of the energy should be negative if $V>0$ and
{\it vice versa}. It is not difficult to show that the
inequalities (23) and (25) have the solutions when $\Delta$ and
$V$ differ from zero, $\Delta>|V|$, and
$k_{y}^2<\Delta^2(\Delta^2-V^2)/(\hbar\upsilon_{F}V)^2$.

As a result, at $V>0$ interface states can be realized exclusively
inside the hole miniband, while the electron miniband consists of
oscillating states only, whose wave functions oscillate, at least,
within $0<x<a$. In Fig.5(a) the hole miniband of the SL with
$a=l/2=30$ nm is plotted for $V=11$ meV, $\Delta=22$ meV, and
$k_{y}l=1$. Upper narrow slice of the valence miniband depicted by
solid line is formed by the interface states. Indeed, it is
possible to verify that any energy value from this range obey the
inequalities (23) and (25) for the above parameters. Dashed (red)
line in the figure shows the interface energy level
$|E|=\hbar\upsilon_{F}|k_{y}|(1-V^2/\Delta^2)^{1/2}$ for an
isolated heterojunction.\cite{RS}

Fig.5(b) represents the squared absolute values of the hole wave
functions corresponding to the valence miniband whose profile is
shown in Fig.5(a) at $k_yl=1$. We plot the probability density for
some interface state (curve 1) with the energy belonging to the
top slice of the valence miniband (solid line in panel (a)). It is
seen that $|\Psi(x)|^2$ exponentially drops towards the interface
$x=a$. For comparison, in Fig.5(b) we plot also the probability
densities for two other states from the valence miniband (curves 2
and 3) oscillating inside the gapless region $0<x<a$. Solving the
Dirac equation at $k_y=0$ it is not difficult to show that
$|\Psi(x)|^2$ is always constant within the gapless region (curve 3).
Note also, that in this case the probability density must be
symmetric with respect to the points $x=a/2$ and
$x=(l+a)/2$, as shown in the figure.

\section{Concluding remarks}

We have considered the simple model of a one-dimensional SL in
which the gap and potential profile are piecewise constant
functions. In the framework of this model the dispersion relation
for Dirac electrons was obtained, and the structure of low-energy
minibands was investigated depending on the potential $V$ and
other parameters of the SL.

It was found that beginning with some critical value $V_c$ of the
relative band shift $V$ the new contact or cone-like Dirac points
appear in the SL spectrum. As a result, at $V>V_c$ the SL becomes
gapless. The contact point exists only at certain $V=V_n$ but
always at $k=k_y=0$, while the cone-like points are situated
symmetrically related to $k_y=0$ at some finite $k_y=\pm k_{yn}$
when $V>V_c$. In the case where the widths of the gapless and
gapped graphene strips in the SL are equal, we found the positions
of the Dirac points in ${\bf k}$-space and obtained an expression
for the threshold potential value $V_c$ corresponding to their appearance. The dispersion relation and carrier velocities were analyzed in the vicinity of the contact and cone-like Dirac points.

Appearance of new Dirac points at $k_y\neq 0$ is typical as well
for SL induced by the potential modulation of a gapless
graphene.\cite{Ar,Vas2,Chiu,Yang,Fertig} Thus, one may conclude
that such a reconstruction of the electron spectrum in a certain
extent represents a universal property of graphene-based
superlattices.

It should be noted that the description of the SL energy spectrum
in our work was carried out within the framework of a one-electron
picture. Recent theoretical\cite{C1,C2,C6} and
experimental\cite{C3,C4,C5} investigations of many-particle
problem in graphene show that one of the most important
consequences of the electron-electron interaction can be the Fermi
velocity renormalization. Such a renormalization, presumably,
results in different values of $\upsilon_F$ in gapless\cite{C6}
and gapped\cite{C7} graphene fractions, which should, of course,
modify the basic dispersion relation [Eq.(9)]. Nevertheless, there
are no yet any rigorous quantitative estimations of the
renormalization effect, and experimentally obtained value of
$\upsilon_F$ is usually close to $10^8$ cm/s. Here, we performed
our calculations assuming the difference in $\upsilon_F$-values in
both graphene fractions to be negligibly small.

Finally, we found that the interface states can exist in the
gap-induced SL at certain conditions.

\section{Acknowledgments}

This work was supported by the Russian Foundation for Basic
Research (Grants No 11-02-00960 and 13-02-00784) and Russian Ministry of
Education and Science (Contract No 07.514.11.4147).


\begin{thebibliography}{99}
\bibitem{text}
The author of Ref.29 asserts that the minigap can
several times exceed the gap value $2\Delta$.
\end{thebibliography}
\end{document}